\documentclass{article}
\usepackage{spconf,amsmath,graphicx}
\usepackage{float} 
\usepackage{enumitem}  
\usepackage{amssymb}  
\setlist{nosep, leftmargin=14pt}
\usepackage{breqn}
\usepackage{mwe} 

\usepackage{fancyhdr}
\usepackage{lipsum} 

\title{TD-NET: A TRI-DOMAIN NETWORK FOR SPARSE-VIEW CT RECONSTRUCTION}
\name{Xinyuan Wang$^{\star}$ \qquad Changqing Su$^{\dagger}$ \qquad Bo Xiong$^{\dagger}$}

\address{$^{\star}$ University College London \\
     Department of Computer Science \\
     London, WC1E 6BT, the United Kingdom \\
     $^{\dagger}$ Peking University \\
     National Engineering Research Center of Visual Technology \\
     Beijing, 100871, China}

\begin{document}
\maketitle
%
\begin{abstract}
Sparse-view CT reconstruction, aimed at reducing X-ray radiation risks, frequently suffers from image quality degradation, manifested as noise and artifacts. Existing post-processing and dual-domain techniques, although effective in radiation reduction, often lead to over-smoothed results, compromising diagnostic clarity. Addressing this, we introduce TD-Net, a pioneering tri-domain approach that unifies sinogram, image, and frequency domain optimizations. By incorporating Frequency Supervision Module(FSM), TD-Net adeptly preserves intricate details, overcoming the prevalent over-smoothing issue. Extensive evaluations demonstrate TD-Net's superior performance in reconstructing high-quality CT images from sparse views, efficiently balancing radiation safety and image fidelity. The enhanced capabilities of TD-Net in varied noise scenarios highlight its potential as a breakthrough in medical imaging.
\end{abstract}
\section{Introduction}
\label{sec:intro}
CT's widespread use in healthcare and other fields \cite{national2011reduced} is tempered by health risks from X-ray radiation. Sparse-view CT, reducing radiation with fewer projections, aims to mitigate this while maintaining diagnostic quality. However, it introduces artifacts due to limited sampling angles, affecting diagnostic accuracy. Traditional methods like FBP \cite{natterer2001mathematics} and ART \cite{natterer2001mathematics}, along with compute-intensive approaches like Total Variation \cite{RUDIN1992259} and Wavelet-based methods \cite{bhatia1996wavelet}, often inadequately address these issues or result in over-smoothed images. Recent deep learning advances, from FBPConvNet \cite{jin2017deep} to DuDoNet \cite{dudonet} and DuDoTrans \cite{dudotrans}, have improved reconstruction but still risk over-smoothness. Our work seeks an alternative that balances detail preservation and smoothness.
In response, We propose the Tri-Domain Network (TD-Net), a novel framework for enhancing CT reconstruction by operating across sinogram, image, and frequency domains. TD-Net addresses the challenge of under-sampling in sparse-view CT, which often leads to artifacts disrupting frequency distribution. By incorporating Frequency Supervision Module(FSM), our framework effectively mitigates these artifacts, resulting in superior reconstruction quality.
The main contributions of TD-Net are:
\begin{itemize}
    \item An end-to-end tri-domain network uniquely crafted for Sparse-view CT, offering a comprehensive solution to image reconstruction challenges.
    \item An architecture that integrates frequency domain feedback into the deep feature extraction process. This novel integration serves as a check against oversmoothing, preserving essential details and sharpness often lost in dual domain reconstruction methods.
\end{itemize}
\section{Methods}
\begin{figure*}[ht!]
    \centering
    \includegraphics[width=\textwidth]{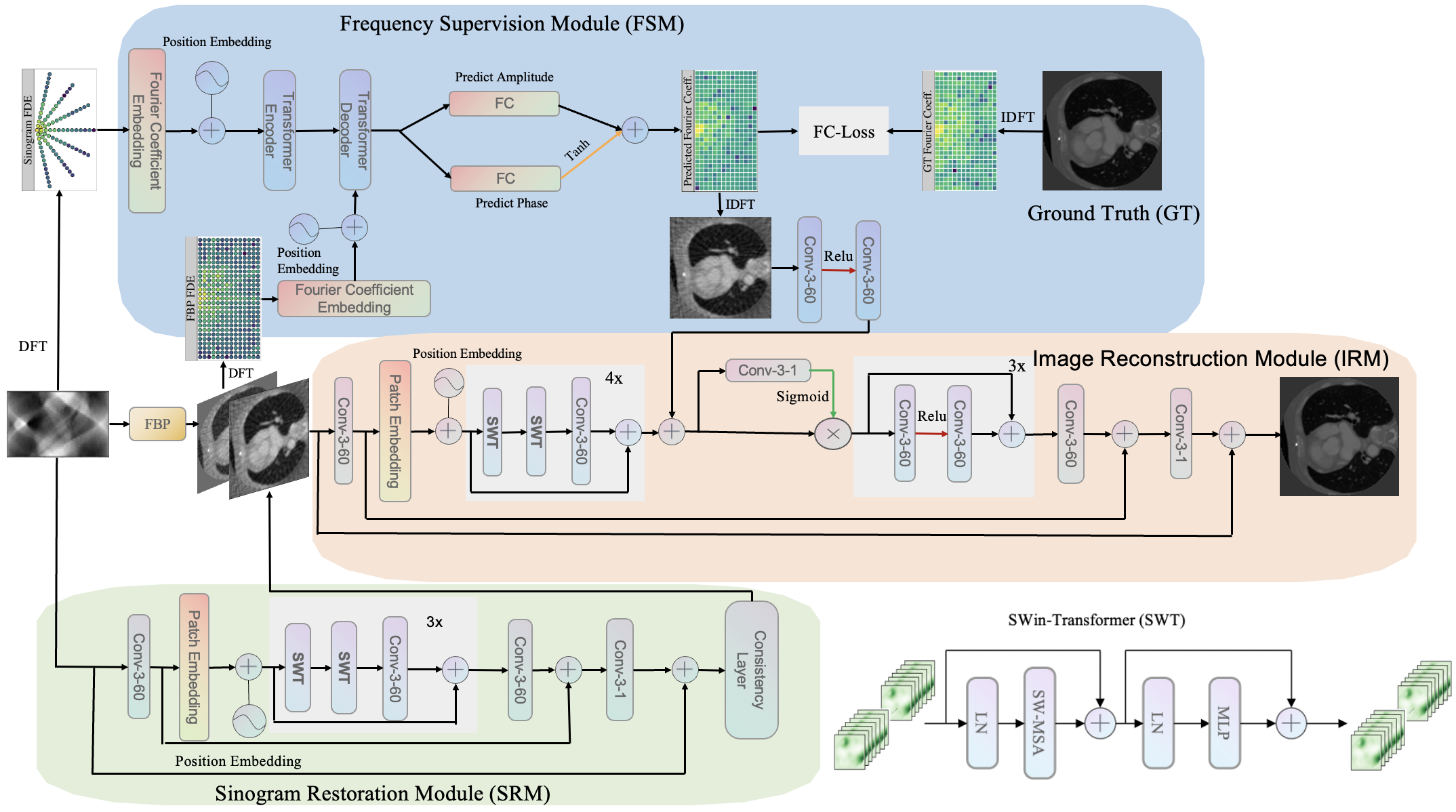}
    \caption{The architecture of our proposed method TD-Net}
    \label{TD-Net}
\end{figure*}

In our proposed architecture, illustrated in Figure \ref{TD-Net} , we integrate three key modules for sparse-view CT reconstruction: (a) Sinogram Restoration Module (SRM), (b) Frequency Supervision Module (FSM), and (c) Image Reconstruction Module (IRM). Consider a sparse-view sinogram represented as $S \in \mathbb{R}^{N \times M}$, where $N$ and $M$ denote view number and sensor number repectively. Filtered Back Projection (FBP) is utilized to reconstruct a preliminary low-quality image $\hat{I}_1$ from $S$.
Concurrently, the SRM module processes $S$, producing an enhanced sinogram $\hat{S}$, and the Consistency Layer generates another image estimate $\hat{I}_2$. Both these low-quality images $\hat{I}_1$ and $\hat{I}_2$ are concatenated and fed into the IRM to predict the CT image $\hat{I}$, which is then supervised with the corresponding clean CT image $I_{\mathrm{GT}} \in \mathbb{R}^{H \times W}$, where $H$ and $W$ represent the height and width of the CT image, respectively.
Furthermore, $S$ and $\hat{I}_1$, along with $I_{\mathrm{GT}}$, are transformed into the Fourier domain. The FSM encodes $S$ and $\hat{I}_1$ into an augmented frequency domain representation $F_{\text {enh }}$, which is then inverse transformed back to the image domain, yielding $\hat{I}_{3}$.
This image $\hat{I}_{3}$ is processed through a two-layer convolution block and then are integrated into the IRM. This integration supervises the image domain to prevent over-smoothing in the final CT reconstruction.
In line with our proposed methodology, the total loss function is defined as a weighted sum of individual loss components.  Consequently, the total loss \(\mathcal{L}_{\text{total}}\) is expressed as follows:
\begin{equation}
\mathcal{L}_{\text{total}} = \lambda_1\mathcal{L}_{\text{SRM}} + \lambda_2\mathcal{L}_{\text{consis}} + \lambda_3\mathcal{L}_{\text{FSM}} + \lambda_4\mathcal{L}_{\text{IRM}}
\end{equation}
For simplicity and based on empirical evaluations, we set all the weight coefficients (\(\lambda_1\), \(\lambda_2\), \(\lambda_3\), and \(\lambda_4\)) to 1.
\subsection{Sinogram Restoration Module (SRM)}
CT imaging's sinogram restoration, key for outlining human anatomy, faces angle-based data overlap. Our Sinogram Restoration Module (SRM) \cite{dudotrans}, using Swin Transformer, overcomes this and traditional CNNs' global processing limits, enhancing efficiency compared to classic Vision Transformers.\\
\\
\textbf{Sinogram Restoration Architecture: }
SRM is composed of multiple consecutive residual blocks, each containing several Swin Transformer modules. This design allows for localized self-attention mechanisms within predefined windows, reducing computational complexity while preserving the effectiveness of global attention. The mathematical formulation of SRM is as follows:
\begin{equation}
\mathbf{F}_{\mathrm{SRM}_i}=\operatorname{Conv}\left(\operatorname{SwinTrans}^{(1: n)}\left(\mathbf{F}_{\mathrm{SRM}_{i-1}}\right)\right)+\mathbf{F}_{\mathrm{SRM}_{i-1}},
\end{equation}
where Conv represents a convolutional layer, and SwinTrans ${ }^{(1: n)}$ signifies a sequence of $n$ Swin Transformer layers applied successively.
The enhanced sinogram $\hat{S}$ is estimated by:
\begin{equation}
\hat{S}=S+\operatorname{Conv}\left(\operatorname{Conv}\left(\mathbf{F}_{\mathrm{SRM}_m}\right)+\mathbf{F}_{\mathrm{SRM}_0}\right)
\end{equation}
with $S$ being the original sinogram.For the input sinogram $\hat{S}$ from SRM, the Consistency Layer $D_{\text {consis }}$ produces an alternate sinogram $\hat{I}_2= D_{\text {Consis }}(\hat{S})$, aligning it with the initial FBP reconstruction $\hat{I}_1$. This alignment aids in synergistic image reconstruction in the RIM module.
\\
\\
\textbf{Loss Function: }
The SRM is supervised using the loss function $\mathcal{L}_{\mathrm{SRM}}$, defined as
$
\mathcal{L}_{\mathrm{SRM}}=\left\|\hat{S}-S_{\mathrm{gt}}\right\|_2
$
where $S_{\mathrm{gt}}$ is the ground truth sinogram, provided during training. This loss function ensures that the restored sinogram closely matches the ground truth, both in terms of global structure and local details.
Also, to maintain consistency with the ground-truth CT image $I_{\mathrm{GT}}$, we introduce a consistency loss function, $\mathcal{L}_{\mathrm{consis}}$, defined as
$
\mathcal{L}_{\text {consis }}=\left\|\hat{I}_2-I_{\mathrm{gt}}\right\|_2
$

\subsection{Frequency Supervision Module (FSM)}
Our Frequency Supervision Module (FSM) utilizes Filtered Back Projection (FBP) based on the Fourier Slice Theorem to address gaps in the Fourier domain caused by sparse-view CT's fewer projections. It fills missing coefficients, reducing artifacts and preserving details. Additionally, FSM aids in image domain reconstruction, ensuring a balance between artifact reduction and detail preservation, crucial for accurate diagnosis\\
\\
\textbf{Fourier Domain Encodings (FDEs):}
For a given sinogram $S$, Fourier Domain Encoding (FDE)\cite{FIT} is computed by applying a Discrete Fourier Transform (DFT), denoted as $F_S = \mathcal{F}(S)$. This results in a complex-valued Fourier spectrum $F_S$. Due to the radial symmetry of $F_S$ (since $S$ is real-valued), we utilize only half of the spectrum, denoted as $F_{S_h}$. This half-spectrum is then transformed into a one-dimensional sequence $S_{\text{enc}} = \text{unroll}\left(F_{S_h}\right)$, consisting of complex Fourier coefficients $\left[s_1, s_2, \ldots, s_N\right]^T$.
The coefficients $s_i$ are normalized into amplitudes $a_i$ and phases $\phi_i$, resulting in a matrix $\mathbf{C} \in \mathbb{R}^{N \times 2}$. We transform each $\left(a_i, \phi_i\right)$ into an $F$-dimensional vector, integrating with positional encoding corresponding to their original polar coordinates in $F_{S_h}$. The final FDE $\boldsymbol{E} \in \mathbb{R}^{N \times F}$ maintains the sequential property, reflecting different resolution levels of $S$.\\
\\
\textbf{Fourier Image Transformer (FIT) Architecture:}
In the FIT framework, we process the output sequence $\mathbf{Z}=\left[z_1, \ldots, z_k\right]$ in $\mathbb{R}^F$ to deduce Fourier amplitudes $\hat{a}_i$ and phases $\hat{\phi}_i$. A significant advancement is integrating these coefficients with 2D positional encodings, enhancing spatial recognition in transformers \cite{wang2021translating}. Employing fast autoregressive transformers for sequential dependencies \cite{katharopoulos2020transformers}, the architecture concludes with an inverse DFT, reconstructing images from Fourier coefficients. For a detailed exposition, see \cite{FIT}.
\\
\\
\textbf{Loss Function: }
The training of FIT involves a specialized loss function, $\mathcal{L}_{\mathcal{F C}}$, combining amplitude loss $\mathcal{L}_{\mathrm{amp}}$ and phase loss $\mathcal{L}_{\text {phase. }}$. Amplitude loss is computed as $\mathcal{L}_{\text {amp }}\left(\hat{a}_i, a_i\right)=1+\left(\hat{a}_i-a_i\right)^2$, and phase loss as $\mathcal{L}_{\text {phase }}\left(\hat{\phi}_i, \phi_i\right) \\ =2-\cos \left(\hat{\phi}_i-\phi_i\right)$. The total Fourier coefficient loss is defined as $
\mathcal{L}_{\mathcal{F C}}(\hat{\mathbf{C}}, \mathbf{C})=\frac{1}{N} \sum_{i=1}^N\left[\mathcal{L}_{\text {amp }}\left(\hat{a}_i, a_i\right) \cdot \mathcal{L}_{\text {phase }}\left(\hat{\phi}_i, \phi_i\right)\right]
$
Then, following the Fourier coefficient loss computation, we apply an Inverse Discrete Fourier Transform (IDFT) to transform the coefficients back to the spatial domain. The output of this process is a reconstructed image $\hat{I}_3$. To ensure fidelity in the spatial domain, we introduce a spatial domain loss $\mathcal{L}_{\text{spatial}}$, which measures the discrepancy between the reconstructed image $\hat{I}_3$ and the ground truth image $I_3$.
The spatial domain loss is defined as:
$\mathcal{L}_{\text{spatial}}(\hat{I}_3, I_{\mathrm{gt}}) =  \left\| \hat{I}_3 - I_{\mathrm{gt}} \right\|_2$ Finally, the overall loss function for the FSM is a weighted sum of the Fourier coefficient loss and the spatial domain loss:
\begin{equation}
\mathcal{L}_{\text{FSM}} = \lambda_{\text{FC}} \cdot \mathcal{L}_{\mathcal{F C}}(\hat{\mathbf{C}}, \mathbf{C}) + \lambda_{\text{spatial}} \cdot \mathcal{L}_{\text{spatial}}(\hat{I}_3, I_{\mathrm{gt}})
\end{equation}
where $\lambda_{\text{FC}}$ and $\lambda_{\text{spatial}}$ are hyperparameters that balance the contribution of each loss component. Based on empirical experience, both $\lambda_{\text{FC}}$ and $\lambda_{\text{spatial}}$ are set to 1.

\subsection{Image Reconstruction Module (IRM)}
The Image Reconstruction Module (IRM) in our framework combines the FBP reconstruction $\hat{I}_1$, Consistency Layer $\hat{I}_2$, and frequency-supervised image $\hat{I}_3$ for CT image synthesis. Feature extraction for $\hat{I}_1$ and $\hat{I}_2$ is conducted using a Swin-transformer, and $\hat{I}_3$ via a dual-layer CNN. The amalgamation of these outputs, post-refinement with an attention layer and residual blocks, undergoes further processing with convolutional layers to enhance details and reduce oversmoothing in the final reconstruction.
\\
\\
\textbf{Loss Function}
The IRM is optimized using a dedicated loss function, $\mathcal{L}_{\text {IRM }}$, which is a composite of multiple terms. This loss function is designed to minimize the discrepancy between the reconstructed image $\hat{I}$ and the ground truth $I_{\mathrm{GT}}$, focusing on both global structural fidelity and local detail preservation. The formulation of the loss function is defined as
$\mathcal{L}_{\mathrm{IRM}} = \left\|\hat{I}-I_{\mathrm{GT}}\right\|_2$

\section{Experimental Results}
\begin{figure}[H]
    \centering
    \includegraphics[width=\columnwidth]{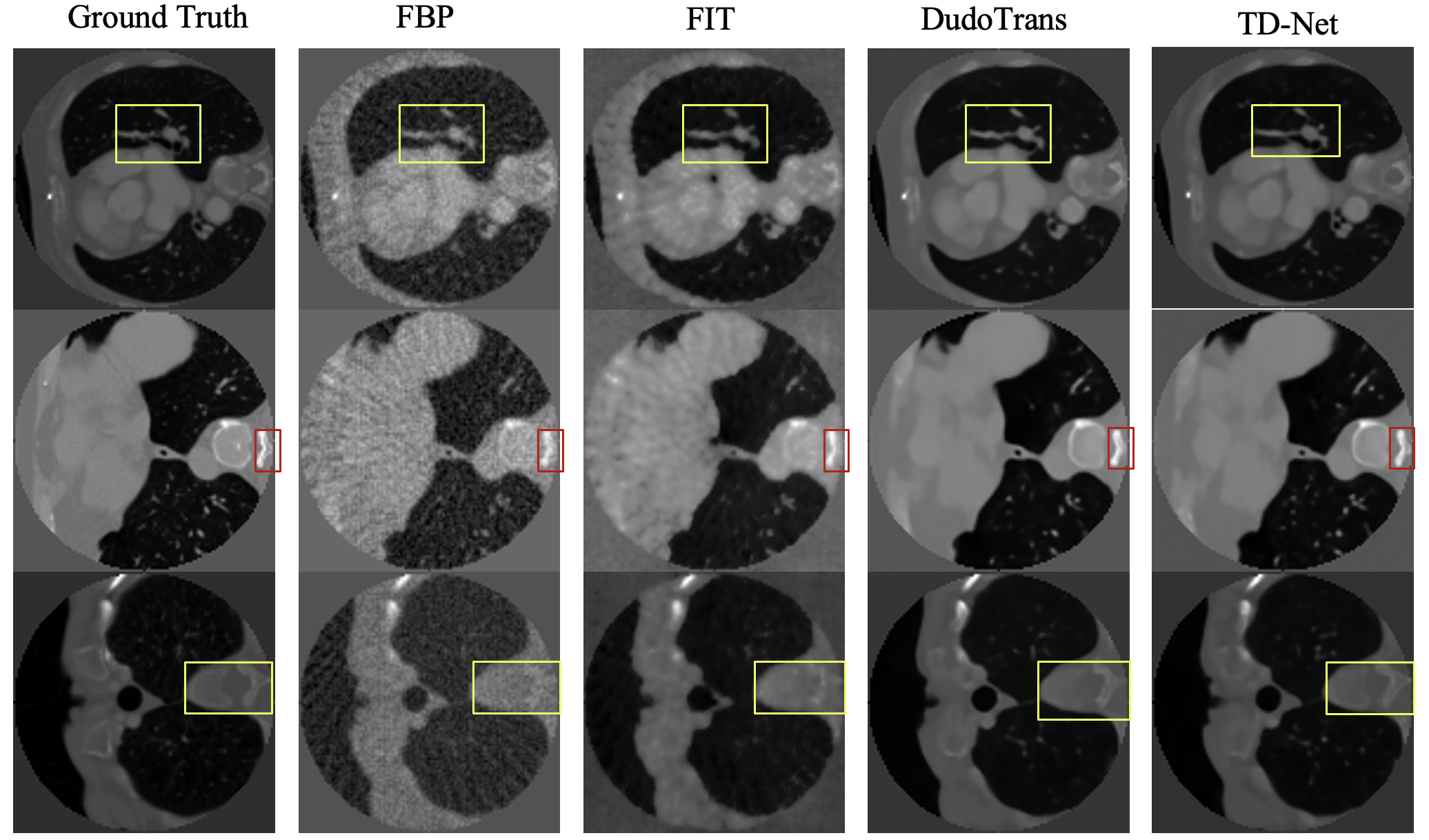}
    \caption{Visual comparisons between different Methods(P=33)}
    \label{allmethods}
\end{figure}

Our study utilized the LoDoPaB dataset \cite{leuschner2019lodopab} for CT reconstruction, selecting 4,000 training, 400 validation, and 3,553 test images, resized to $111 \times 111$ pixels with $P=21, 33, 45$ projections. To prevent inverse crime \cite{wirgin2004inverse}, images were upscaled and injected with $5\%$ Gaussian noise before sinogram generation. Our PyTorch \cite{paszke2019pytorch} framework used RAdam optimizer, learning rate 0.0001, weight decay 0.01, and ReduceLROnPlateau scheduler. Training was conducted for 200 epochs on an Nvidia 3080Ti GPU, using validation MSE for learning rate adjustments.
\begin{table}[ht]
\renewcommand{\arraystretch}{0.8} 
\caption{Performance metrics of different methods at varying angles}
\label{tab:metrics_comparison}
\centering
\begin{tabular}{|c|c|c|c|}
\hline
\textbf{Method} & \textbf{PSNR (dB)} & \textbf{SSIM} & \textbf{RMSE} \\
\hline
FBP (P=21) & 23.098 & 0.504 & 0.041  \\
FIT (P=21) \cite{FIT}& 28.235 & 0.756  & 0.021 \\
DudoTrans (P=21) \cite{dudotrans} & 32.012 & 0.869 & 0.014 \\
TD-Net(ours) (P=21) & \textbf{32.327} & \textbf{0.875} & \textbf{0.013} \\
\hline
FBP (P=33) & 24.67 & 0.582 & 0.037 \\
FIT (P=33)\cite{FIT}& 29.961 & 0.805  & 0.0174 \\
DudoTrans (P=33) \cite{dudotrans} & 33.747 & 0.899 & 0.011 \\
TD-Net(ours)(P=33) & \textbf{33.981} & \textbf{0.903} & \textbf{0.010} \\
\hline
FBP (P=45) & 25.497 & 0.633 & 0.035 \\
FIT (P=45) \cite{FIT}& 31.086 & 0.833 & 0.015 \\
DudoTrans (P=45) \cite{dudotrans} & 34.760 & 0.915 & 0.010 \\
TD-Net(ours) & \textbf{34.925} & \textbf{0.917} & \textbf{0.010} \\
\hline
\end{tabular}
\end{table}
The table \ref{tab:metrics_comparison} clearly indicate that TD-Net outperforms the recognized state-of-the-art DuDoTrans. Visual analysis from the figure \ref{allmethods} further confirms TD-Net's enhanced ability to restore details while effectively avoiding the oversmoothing prevalent in competing approaches.
\section{Conclusion}
Our tri-domain approach merges frequency and spatial domains, outperforming the state-of-the-art DuDoTrans as confirmed by our experiments. This underscores the frequency domain's role in improving image reconstruction and reducing oversmoothing.

\section{Acknowledgments}
 This work was supported by National Natural Science Foundation of China(Grant No. 62371006). There is no other conflicts of interest.

"\section{COMPLIANCE WITH ETHICAL STANDARDS}
This retrospective study used openly accessible images, negating the need for additional approval.
\bibliographystyle{IEEEbib}
\bibliography{refs}

\end{document}